\documentclass[twocolumn, showpacs, nofootinbib, aps, prd]{revtex4}

\usepackage{graphicx}
\usepackage{dcolumn}
\usepackage{bm}
\usepackage{hyperref}
\usepackage{color}
\usepackage{amsfonts}

\begin{document}

\title{Uncertainties of the $B\to D$ transition form factor from the $D$-Meson Leading-Twist Distribution Amplitude}

\author{Yi Zhang$^{1}$}
\author{Tao Zhong$^{1}$}
\email{zhongtao@htu.edu.cn}
\author{Xing-Gang Wu$^{2}$}
\email{wuxg@cqu.edu.cn}
\author{Ke Li$^{1}$}
\author{Hai-Bing Fu$^{3}$}
\author{Tao Huang$^4$}
\email{huangtao@ihep.ac.cn}

\address{$^1$ College of Physics and Materials Science, Henan Normal University, Xinxiang 453007, People's Republic of China\\
$^2$ Department of Physics, Chongqing University, Chongqing 401331, People's Republic of China \\
$^3$ School of Science, Guizhou Minzu University, Guiyang 550025, People's Republic of China \\
$^4$ Institute of High Energy Physics and Theoretical Physics Center for Science Facilities, Chinese Academy of Sciences, Beijing 100049, People's Republic of China}

\date{\today}

\begin{abstract}
The $B\to D$ transition form factor (TFF) $f^{B\to D}_+(q^2)$ is determined mainly by the $D$-meson leading-twist distribution amplitude (DA) , $\phi_{2;D}$, if the proper chiral current correlation function is adopted within the light-cone QCD sum rules. It is therefore significant to make a comprehensive study of DA $\phi_{2;D}$ and its impact on $f^{B\to D}_+(q^2)$. In this paper, we calculate the moments of  $\phi_{2;D}$ with the QCD sum rules under the framework of the background field theory. New sum rules for the leading-twist DA moments $\left<\xi^n\right>_D$ up to fourth order and up to dimension-six condensates are presented.  At the scale $\mu = 2 \rm GeV$, the values of the first four moments are: $\left<\xi^1\right>_D = -0.418^{+0.021}_{-0.022}$, $\left<\xi^2\right>_D = 0.289^{+0.023}_{-0.022}$, $\left<\xi^3\right>_D = -0.178 \pm 0.010$ and $\left<\xi^4\right>_D = 0.142^{+0.013}_{-0.012}$. Basing on the values of $\left<\xi^n\right>_D(n=1,2,3,4)$, a better model of $\phi_{2;D}$ is constructed. Applying this model for the TFF $f^{B\to D}_+(q^2)$ under the light cone sum rules, we obtain $f^{B\to D}_+(0) = 0.673^{+0.038}_{-0.041}$ and $f^{B\to D}_+(q^2_{\rm max}) = 1.124^{+0.053}_{-0.058}$. The uncertainty of $f^{B\to D}_+(q^2)$ from $\phi_{2;D}$ is estimated and we find its impact should be taken into account, especially in low and central energy region. The branching ratio $\mathcal{B}(B\to Dl\bar{\nu}_l)$ is calculated, which is consistent with experimental data.

\end{abstract}


\maketitle

\section{introduction}

The $B\to D^{(\ast)}$ decays have received a lot of attention in recent years. Experimentally, the BABAR Collaboration measured the semi-leptonic decays $B\to D^{(\ast)}l\bar{\nu}_l$ in 2012 \cite{BABAR_Aubert:2007dsa, BABAR_Lees:2012xj, BABAR_Lees:2013uzd}, and these decays were also measured by the Belle \cite{BELLE_Bozek:2010xy, BELLE_Huschle:2015rga, BELLE_Matyja:2007kt} and LHCb Collaborations \cite{LHCB_Aaij:2015yra} in 2015. Theoretically, the semi-leptonic decays $B\to D^{(\ast)}l\bar{\nu}_l$ are studied by the heavy quark effective theory (HQET) \cite{HQET_Fajfer:2012vx}, perturbative QCD (pQCD) factorization approach \cite{PQCD_Fan:2013qz, PQCD_Fan:2015kna}, light-cone sum rules (LCSR) \cite{LCSR_Zuo:2006dk, Zuo:2006re, Fu:2013wqa, Wang:2017jow}, and the lattice QCD theory \cite{LQCD_Lattice:2015rga, LQCD_Na:2015kha} within the framework of Standard Model (SM) and the new physics model \cite{NP_Celis:2012dk, NP_Celis:2013jha, NP_Li:2016vvp}.

The $D$-meson distribution amplitude (DA) is an important input for theoretical studies. By using the usual correlation function in the LCSR calculation, the $B\to D$ transition form factor (TFF) $f^{B\to D}_+(q^2)$ is represented as a complex formula containing the $D$-meson twist-$2,3,\cdots$ DAs. If a proper chiral current correlation function is adopted, the TFF $f^{B\to D}_+(q^2)$ shall be dominated by the contribution of $\phi_{2;D}$ \cite{LCSR_Zuo:2006dk, Zuo:2006re, Fu:2013wqa},
\begin{eqnarray}
f_+^{B\to D}(q^2) &=& \frac{m_b^2 f_D}{m_B^2 f_B} e^{m_B^2/M^2} \int^1_\Delta \frac{du}{u} \phi_{2;D}(u) \nonumber\\
&& \times\exp \left[ -\frac{m_b^2 - \bar{u} (q^2 - u m_D^2)}{u M^2} \right],
\label{f+_LCSR}
\end{eqnarray}
where $m_b$ is the $b$-quark mass, $m_{B(D)}$ and $f_{B(D)}$ are the $B(D)$-meson mass and decay constant, $s^B_0$ is threshold parameter, $M$ is the Borel parameter, and the lower limit of integral takes the form
\begin{eqnarray}
\Delta &=& \frac{1}{2m_D^2} \left[ \sqrt{(s_0^B - q^2 - m_D^2)^2 + 4m_D^2 (m_b^2 - q^2)} \right. \nonumber\\
&& \left. -(s_0^B - q^2 - m_D^2) \right].  \nonumber
\end{eqnarray}
Eq.(\ref{f+_LCSR}) reduces the error sources of $f^{B\to D}_+(q^2)$, such as the uncertain twist-3 DAs disappear in the LCSR. In turn, it provides us with a precise platform for testing the behavior of the leading-twist DA $\phi_{2;D}$~\cite{Huang:2013yya}.

In the existing researches on $f^{B\to D}_+(q^2)$, the DA $\phi_{2;D}$ is simply treated as an input parameter, whose error is often not considered. The most simple model is based on the expansion of the Gegenbauer polynomials, which reads \cite{MODELI_Kurimoto:2002sb, MODELI_Keum:2003js}:
\begin{eqnarray}
\phi_{2;D}^{\rm KLS}(x) = 6x(1-x)\left[ 1 + C_D(1-2x) \right],
\label{modelI}
\end{eqnarray}
where $x$ stands for the momentum fraction of the light quark and the shape parameter $C_D$ is usually taken as $\sim 0.7$, corresponding to a peak around $x \sim 0.3$. Considering a simple harmonic-like $k_\perp$-dependence in the $D$-wavefunction, its DA is improved as \cite{MODELII_Li:2008ts}:
\begin{eqnarray}
\phi_{2;D}^{\rm LLZ}(x) = 6x(1-x)\left[ 1 + C_D(1-2x) \right] \exp \left[ -\frac{\omega^2 b^2}{2} \right],
\label{modelII}
\end{eqnarray}
where the parameters $b=0.38{\rm GeV}^{-1}$, $C_D=0.5$, $\omega=0.1{\rm GeV}$ \cite{MODELII_Li:2008ts}. By employing the solution of a relativistic scalar harmonic oscillator potential \cite{DrozVincent:1978yk, Sazdjian:1985pg} for the orbital part of the wavefunction \cite{Bauer:1988fx, Wirbel:1985ji}, the authors of Ref.\cite{MODELIII_Li:1999kna} suggest a Gaussian-type model:
\begin{eqnarray}
\phi_{2;D}^{\rm LM}(x) = N_D \sqrt{x(1-x)} \exp \left[ -\frac{1}{2}\left( \frac{x m_D}{\omega} \right)^2 \right],
\label{modelIII}
\end{eqnarray}
where $m_D=1.87{\rm GeV}$, $N_D=4.86952$, $f_D=220{\rm MeV}$ and $\omega=0.8{\rm GeV}$. By using the Brodsky-Huang-Lepage prescription \cite{BHL1, BHL2, BHL3}, Ref.\cite{MOLELIV_Guo:1991eb} proposes a light-cone harmonic oscillator model:
\begin{eqnarray}
\phi_{2;D}^{\rm GH}(x) = N_D x(1-x) \exp \left[ -b_D^2 \frac{\hat{m}_c^2 x + \hat{m}_d^2 (1-x)}{x(1-x)} \right],
\label{modelIV}
\end{eqnarray}
where the constituent quark masses $\hat{m}_c=1.3{\rm GeV}$ and $\hat{m}_d=0.35{\rm GeV}$, $N_D=19.908$, and $b_D^2=0.292{\rm GeV}^{-2}$ \cite{LCSR_Zuo:2006dk}. By including the Melosh rotation effect into the spin space, a more complete form than the model (\ref{modelIV}) has also been presented in Ref.\cite{MOLELIV_Guo:1991eb}. In addition, there are other two $D$-meson leading-twist DA models, the exponential model \cite{MODELV_Grozin:1996pq} and the one obtained by solving the equations of motion without three-parton contributions \cite{MODELVI_Kawamura:2001jm}.

As a matter of fact, our understanding of $\phi_{2;D}$ is far from enough, a detail analysis on the uncertainty of various DA models is necessary. In this article, we shall improve the $\phi_{2;D}$ model (\ref{modelIV}) to a more accurate form. As we have done in Refs.\cite{BHL_Zhong:2014jla, BHL_Zhong:2014fma, BHL_Zhong:2016kuv}, its input parameters shall be fixed by using several reasonable constraints, such as the probability of finding the leading Fock-state $\left|\bar{c}q\right>$ in the $D$-meson Fock-state expansion, the normalization condition, and the known $\phi_{2;D}$ Gegenbauer moments. Those Gegenbauer moments shall be computed by using the QCD sum rules \cite{SVZ_Shifman:1978bx} in the framework of background field theory (BFT) \cite{BHL_Zhong:2014jla, BFT_Huang:1986wm, BFT_Huang:1989gv}. As a further step, we shall analyze the properties of the model in detail, and the influence of $\phi_{2;D}$ on $f^{B\to D}_+(q^2)$ shall also be presented.

The remaining parts of the paper are organized as follows. An improved model for the $D$-meson leading twist DA $\phi_{2;D}$ is given in Sec.II. Procedures for deriving the QCD sum rules for the moments of $\phi_{2;D}$ in the BFT are given in Sec.III. For convenience, we present the explicit expressions of those moments in the Appendix. Numerical results and discussions are presented in Sec.IV. Sec.V is reserved for a summary.

\section{An improved Model for the $D$-Meson Leading-Twist DA $\phi_{2;D}$}

As discussed in Refs.\cite{BHL_Zhong:2014fma,BHL_Zhong:2016kuv}, we improve the harmonic oscillator model of the $D$-meson leading-twist wavefunction $\Psi_{2;D}(x,\mathbf{k}_\perp)$ suggested in Ref.\cite{LCSR_Zuo:2006dk} as
\begin{eqnarray}
\Psi_{2;D}(x,\mathbf{k}_\perp) = \chi_{2;D}(x,\mathbf{k}_\perp) \Psi_{2;D}^R(x,\mathbf{k}_\perp),
\label{wf}
\end{eqnarray}
where $\mathbf{k}_\perp$ is the transverse momentum, $\chi_{2;D}(x,\mathbf{k}_\perp)$ stands for the spin-space wavefunction and $\Psi_{2;D}^R(x,\mathbf{k}_\perp)$ indicates the spatial wavefunction. The spin-space wavefunction $\chi_{2;D}(x,\mathbf{k}_\perp)$ takes the form \cite{Huang:1994dy}
\begin{eqnarray}
\chi_{2;D}(x,\mathbf{k}_\perp) = \frac{\hat{m}_c x + \hat{m}_q(1-x)}{\sqrt{\mathbf{k}_\perp^2 + \left[ \hat{m}_c x + \hat{m}_q(1-x) \right]^2}},
\label{chiwf}
\end{eqnarray}
where $\hat{m}_c$ and $\hat{m}_q$ are constituent quark masses of the $D$-meson, and we adopt $\hat{m}_c = 1.5 \rm GeV$ and $\hat{m}_q = 0.3 \rm GeV$. $q$  stands for the light quark, $q=u$ is for $\overline{D}^0$ and $q=d$ is for $D^-$. The spatial wavefunction takes the form
\begin{eqnarray}
\Psi_{2;D}^R(x,\mathbf{k}_\perp) &=& A_D \varphi_D(x) \nonumber\\
&\times& \exp \left[ -\frac{1}{\beta_D^2} \left( \frac{\mathbf{k}_\perp^2 + \hat{m}_c^2}{1-x} + \frac{\mathbf{k}_\perp^2 + \hat{m}_q^2}{x} \right) \right],
\label{psirwf}
\end{eqnarray}
where $A_D$ is the normalization constant, $\beta_D$ is the harmonious parameter that dominates the wavefunction's transverse distribution, and $\varphi_D(x)$ dominates the wavefunction's longitudinal distribution, which can be expanded as a Gegenbauer polynomial,
\begin{eqnarray}
\varphi_D(x) = 1 + \sum^4_{n=1} B_n^D C_n^{3/2}(2x-1).
\end{eqnarray}
Using the relationship between the $D$-meson leading-twist wavefunction and its DA,
\begin{eqnarray}
\phi_{2;D}(x,\mu_0) = \frac{2\sqrt{6}}{f_D} \int_{|\mathbf{k}_\perp|^2\leq \mu_0^2} \frac{d^2\mathbf{k}_\perp}{16\pi^3} \Psi_{2;D}(x,\mathbf{k}_\perp),
\end{eqnarray}
we obtain a new model for $\phi_{2;D}$, i.e.
\begin{eqnarray}
\phi_{2;D}(x,\mu_0) &=& \frac{\sqrt{6} A_D \beta_D^2}{\pi^2 f_D} x(1-x) \varphi_D(x) \nonumber\\
&\times& \exp \left[ - \frac{\hat{m}_c^2x + \hat{m}_q^2(1-x)}{8\beta_D^2 x(1-x)} \right] \nonumber\\
&\times& \left\{ 1 - \exp \left[ -\frac{\mu^2_0}{8\beta_D^2 x(1-x)} \right] \right\},
\label{phi}
\end{eqnarray}
where $\mu_0 \sim \Lambda_{\rm QCD}$ is the factorization scale. Because $\hat{m}_c \gg \Lambda_{\rm QCD}$, the spin-space wavefunction $\chi_D \to 1$. In this work we ignore the (constituent) mass difference between $u$ and $d$ quarks, the wavefunction $\Psi_{2;D}(x,\mathbf{k}_\perp)$ and the DA $\phi_{2;D}(x,\mu)$ are the same for both $\overline{D}^0$ and $D^-$. By replacing $x$ with $1-x$ in Eqs.(\ref{wf}, \ref{phi}), one can obtain the leading-twist wavefunction and DA of $D^0$ and $D^+$.

The model parameters $A_D$, $B_n^D$ and $\beta_D$ are scale dependent, their values at an initial scale $\mu_0$ can be determined by reasonable constraints, and their values at any other scale $\mu$ can be obtained via the evolution equation \cite{BHL_Zhong:2014fma, BHL_Zhong:2016kuv}. More explicitly, we shall adopt the following constraints to fix the parameters:
\begin{itemize}
\item The normalization condition,
\begin{eqnarray}
\int^1_0 dx \phi_{2;D}(x,\mu_0) = 1. \label{NC}
\end{eqnarray}

\item The probability of finding the leading Fock-state $\left.\left.\right|\bar{c}q\right>$ in the $D$-meson Fock state expansion,
\begin{eqnarray}
P_D &=& \frac{A_D^2 \beta_D^2}{4\pi^2} x(1-x) \varphi_D^2(x) \nonumber\\
&\times& \exp \left[ -\frac{m_c^2x + m_q^2 (1-x)}{4\beta^2_D x(1-x)} \right].   \label{P}
\end{eqnarray}
We will take $P_D \simeq 0.8$ ~\cite{MOLELIV_Guo:1991eb} in subsequent calculation. Numerically, we find that similar to the case of heavy pseudo-scalar meson \cite{BHL_Zhong:2014fma}, our model depends very little on the value of $P_D$.

\item The Gegenbauer moments of $\phi_{2;D}(x,\mu_0)$ can be calculated by the following way,
\begin{eqnarray}
a_n^D(\mu_0) = \frac{\int^1_0 dx \phi_{2;D}(x,\mu_0)C_n^{3/2}(2x-1)}{\int^1_0 dx 6x(1-x) [C_n^{3/2}(2x-1)]^2}. \label{an}
\end{eqnarray}
If knowing their values, we can inversely determine the behavior of $\phi_{2;D}(x,\mu_0)$.

\end{itemize}

\section{Sum rules of the moments of the leading-twist DA $\phi_{2;D}$}

To derive the sum rules for the $D$-meson leading-twist DA $\phi_{2;D}$, we introduce the following correlation function
\begin{eqnarray}
\Pi^{(n,0)}_D (z,q) && = i \int d^4x e^{iq\cdot x} \left<0\left| T \left\{ J_n(x) J^\dag_0(0) \right\} \right|0\right>
\nonumber\\ &&
= (z\cdot q)^{n+2} I^{(n,0)}_D (q^2) ,
\label{correlator}
\end{eqnarray}
where $z^2 = 0$, $n=0,1,2,\cdots$, and the currents
\begin{eqnarray}
J_n(x) &=& \bar{c}(x) {z\!\!\!\slash} \gamma_5 (i z\cdot \tensor{D})^n q(x), \\
J^\dagger_0(0) &=& \bar{q}(0) {z\!\!\!\slash} \gamma_5 c(0).
\end{eqnarray}
Following the standard procedures of QCD sum rules, we first apply the operator product expansion (OPE) for the correlation function (\ref{correlator}) in the deep Euclidean region. With the basic assumption of BFT and the corresponding Feynman rules, Eq.(\ref{correlator}) can be rewritten as
\begin{eqnarray} &&
\Pi^{(n,0)}_D (z,q) = i \int d^4x e^{iq\cdot x} \nonumber\\ &&
\times \left\{ - {\rm Tr} \left<0\left| S_F^c(0,x) {z\!\!\!\slash} \gamma_5 (i z\cdot \tensor{D})^n S_F^q(x,0) {z\!\!\!\slash} \gamma_5  \right|0\right> \right. \nonumber\\ &&
+ \left. {\rm Tr} \left<0\left| S_F^c(0,x) {z\!\!\!\slash} \gamma_5 (i z\cdot \tensor{D})^n \bar{q}(0)q(x) {z\!\!\!\slash} \gamma_5  \right|0\right> \right\} \nonumber\\&&
+ \cdots,
\label{ope}
\end{eqnarray}
where $S_F^c(0,x)$ and $S_F^q(x,0)$ are quark propagators in the background field, and $(i z\cdot \tensor{D})^n$ stands for the vertex operators. The tedious expressions of the propagators and vertex operators with terms leading to dimension-six condensates in the sum rules can be found in Ref.\cite{BHL_Zhong:2014jla}.

\begin{figure*}
\includegraphics[width=0.85\textwidth]{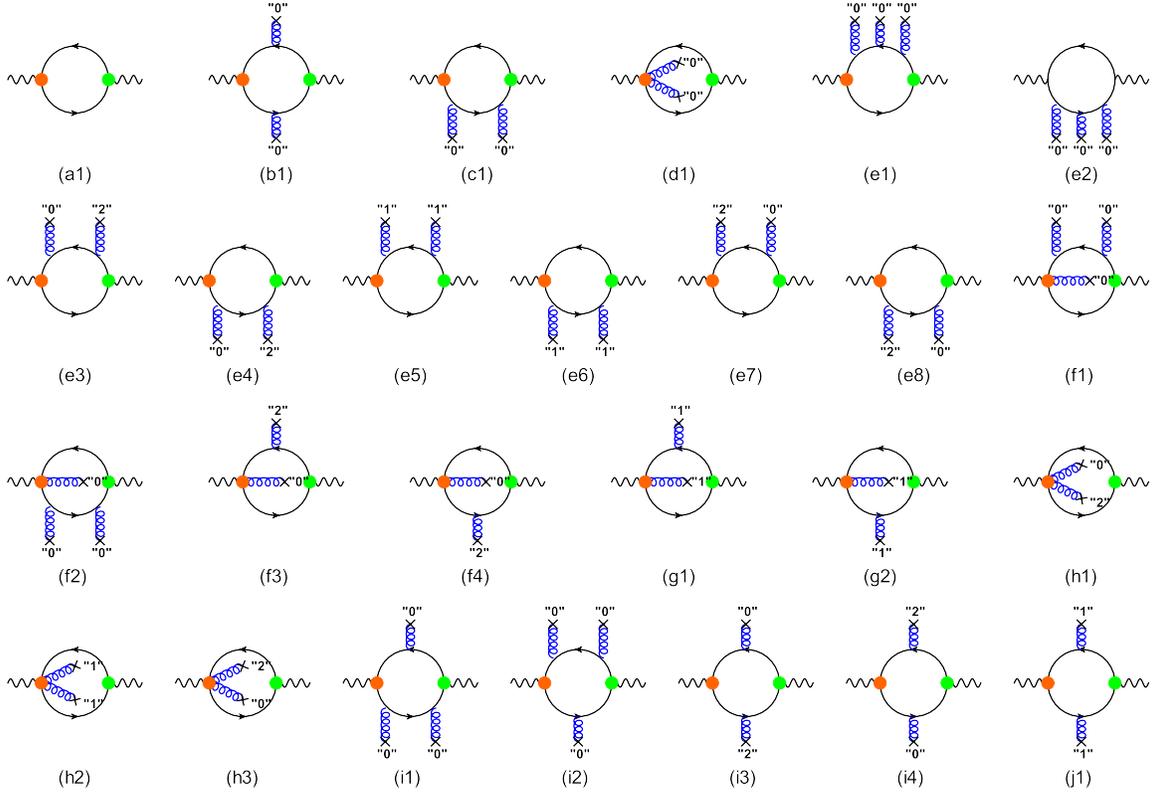}
\caption{Feynman diagrams for the first term of Eq.(\ref{ope}). The left big dot and the right big dot stand for the vertex operators $\not\! z \gamma_5 (z\cdot \tensor{D})^n$ and $\not\! z \gamma_5$ in the currents $J_n(x)$ and $J^\dagger_0(0)$, respectively. The cross symbol attached to the gluon line indicates the tensor of the local gluon background field, and ``$n$" indicates $n_{\rm th}$-order covariant derivative. The diagrams whose contributions vanish in the sum rules are not shown.}
\label{feyna}
\end{figure*}

\begin{figure*}
\includegraphics[width=0.85\textwidth]{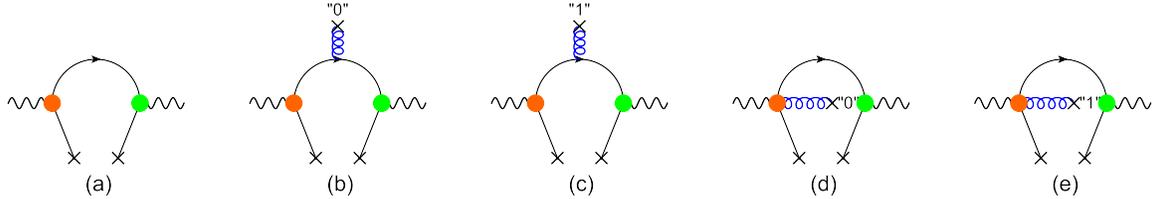}
\caption{Feynman diagrams for the second term of Eq.(\ref{ope}). The left big dot and the right big dot stand for the vertex operators $\not\! z \gamma_5 (z\cdot \tensor{D})^n$ and $\not\! z \gamma_5$ in the currents $J_n(x)$ and $J^\dagger_0(0)$, respectively. The cross symbol attached to the gluon line indicates the tensor of the local gluon background field, and ``$n$" indicates $n_{\rm th}$-order covariant derivative, and the cross symbol attached to the quark line stands for the local light $u$ or $d$ quark background field.}
\label{feynb}
\end{figure*}

Fig.(\ref{feyna}) and Fig.(\ref{feynb})  show the Feynman diagrams for the first and the second terms in Eq.(\ref{ope}), respectively. In those two figure, the left big dot and the right big dot stand for the vertex operators $\not\! z \gamma_5 (z\cdot \tensor{D})^n$ and $\not\! z \gamma_5$ in the currents $J_n(x)$ and $J^\dagger_0(0)$, respectively; the cross symbol attached to the gluon line indicates the tensor of the local gluon background field, and ``$n$" indicates $n_{\rm th}$-order covariant derivative; the cross symbol attached to the quark line stands for the local light $u$ or $d$ quark background field.

Fig.(\ref{feyna}.a1) gives the perturbative contribution, Figs.(\ref{feyna}.b1, \ref{feyna}.c1, \ref{feyna}.d1) give the contributions proportional to dimension-four gluon condensate $\left<\alpha_sG^2\right>$, and the remaining diagrams in Fig.(\ref{feyna}) give the contributions proportional to dimension-six gluon condensate $\left<g_s^3fG^3\right>$. Fig.(\ref{feynb}) gives the terms involving dimension-three quark condensate $\left<\bar{q}q\right>$, dimension-five quark-gluon mixing condensate $\left<g_s\bar{q}\sigma TGq\right>$ and dimension-six quark condensate $\left<g_s\bar{q}q\right>^2$. There is infrared divergence in Figs.(\ref{feyna}.e1, \ref{feyna}.e3, \ref{feyna}.e5, \ref{feyna}.e7, \ref{feyna}.f1, \ref{feyna}.f3, \ref{feyna}.g1, \ref{feyna}.i2, \ref{feyna}.i4, \ref{feyna}.j1), which contain the terms proportional to $\tilde{\Pi}$,
\begin{eqnarray}
\tilde{\Pi} = \mu^{2\epsilon} \int \frac{d^Dp_2}{(2\pi)^D} \frac{(2p_2\cdot z - p_1\cdot z)^n  \times \cdots}{[(q-p_2)^2 - m_c^2]^\alpha (p_2^2)^\beta}, \;\;  (\alpha < \beta)
\end{eqnarray}
where we have completed the integration over $x$, the $c$-quark momenta $p_1$ and $p_{2}$ indicates the $u/d$ quark momentum, and the ellipsis ``$\cdots$'' stands for the possible Lorenz structures, such as $p_2^\mu$, $p_2^\mu p_2^\nu$, and etc. Taking the limit, $m_{u/d}^2 \to 0$, the infrared divergence appears in $\tilde{\Pi}$. We adopt the $D$-dimensional regularization approach to deal with the infrared divergence, $D = 4 - 2\epsilon$ ($\epsilon\to 0$). Then our task is to extract the divergent terms proportional to $1/\epsilon$. Using Feynman parameterization formula,
\begin{eqnarray}
\frac{1}{A^\alpha B^\beta} = \frac{\Gamma(\alpha+\beta)}{\Gamma(\alpha)\Gamma(\beta)} \int^1_0 dx \frac{x^{\alpha-1} (1-x)^{\beta-1}}{[Ax + B(1-x)]^{\alpha+\beta}}
\end{eqnarray}
and completing the integration over the momentum $p_2$, we get the key integration for $\tilde{\Pi}$£º
\begin{eqnarray}
I(m,a,b,c) &=& \int^1_0 dx (2x-1)^m x^{-a-\epsilon} (1-x)^b \nonumber\\
&\times& \left( 1 - \frac{-q^2}{-q^2 + m_c^2}x \right)^{-c-\epsilon},
\label{FPI}
\end{eqnarray}
where $m(\leq n),a,b,c$ are integers. Eq.(\ref{FPI}) can be further represented as
\begin{eqnarray}
I(m,a,b,c) &=& \sum^m_{k=0} \frac{(-1)^k m!}{k! (m-k)!} \int^1_0 dx x^{m-k-a-\epsilon} (1-x)^{k+b} \nonumber\\
&\times& \left( 1 - \frac{-q^2}{-q^2 + m_c^2}x \right)^{-c-\epsilon}.
\label{FPIE} 
\end{eqnarray}
It can be simplified with the help of the hypergeometric function, i.e.
\begin{eqnarray}
F(\alpha,\beta,\gamma,Z) &=& \frac{\Gamma(\gamma)}{\Gamma(\beta)\Gamma(\gamma-\beta)} \nonumber\\
&\times& \int^1_0 dx x^{\beta-1} (1-x)^{\gamma-\beta-1}(1-Z x)^{-\alpha} \nonumber\\
&=& \sum^\infty_{l=0} \frac{\left(\alpha\right)_l \left(\beta\right)_l}{l! \left(\gamma\right)_l} Z^l,
\label{HF} 
\end{eqnarray}
where $\left| Z \right| < 1$ and $\left(\lambda\right)_l = \Gamma(\lambda + l)/\Gamma(\lambda)$, we obtain
\begin{eqnarray}
I(m,a,b,c) &=& \sum^m_{k=0} \frac{(-1)^k m!}{k! (m-k)!} \frac{\Gamma(k+b+1)}{\Gamma(c+\epsilon)} \nonumber\\
&\times& \sum^\infty_{l=0} \frac{\Gamma(l+c+\epsilon) \Gamma(l+m-k-a+1-\epsilon)}{l! \Gamma(l+m-a+b+2-\epsilon)} \nonumber\\
&\times& \left( \frac{-q^2}{-q^2+m_c^2} \right)^l.
\label{FPIS} 
\end{eqnarray}
The infrared divergence appears in $\Gamma(l+m-k-a+1-\epsilon)$ at the lowest several $l$-terms. We adopt the $\overline{\rm MS}$-scheme to deal with the divergent terms, which shall be absorbed into the renormalized $D$-meson leading-twist DA~\cite{BHL_Zhong:2014jla, MS_Li:2012gr}.

On the other hand, the correlation function (\ref{correlator}) can be calculated by inserting a completed set of intermediate hadronic states in the physical region. With the definition
\begin{eqnarray}
\left<0\left| \bar{c}(0) \not\! z \gamma_5 (i z \cdot \tensor{D})^n q(0) \right|D(q)\right>    \nonumber\\
=i(z \cdot q)^{n+1}f_{D} \left<\xi^n\right>_{D},
\label{xinD}
\end{eqnarray}
and the quark-hadron duality, the hadron expression of $\Pi^{(n,0)}_D (z,q) $ can be obtained. In Eq.(\ref{xinD}),
\begin{eqnarray}
\left<\xi^n\right>_{D}=\int^1_0 du(2u-1)^n\phi_{2;D}(u)
\label{xin}
\end{eqnarray}
is the $n_{\rm th}$-order moment of $\phi_{2;D}$. The $0_{\rm th}$-order moment corresponds to the normalization condition for $\phi_{2;D}$,
\begin{eqnarray}
\left<\xi^0\right>_{D}=\int^1_0 du\phi_{2;D}(u)=1.
\end{eqnarray}

The operator expansion of the correlation function (\ref{correlator}) and its hadron expansion in deep Euclidean region can be matched by the dispersion relation. By further applying the Borel transformation for both sides, the sum rules for the moments of the $D$-meson leading-twist DA $\phi_{2;D}$ can be written as
\begin{eqnarray}
\left<\xi^n\right>_D &=& \frac{M^2 e^{\frac{m_D^2}{M^2}}}{f_D^2} \left\{ \frac{1}{\pi} \frac{1}{M^2} \int^{s^D_0}_{t_{min}} ds e^{-\frac{s}{M^2}} {\rm Im} I_{\rm pert}(s) \right. \nonumber\\
&+& \hat{L}_M I_{\left<\bar{q}q\right>}(-q^2) + \hat{L}_M I_{\left<G^2\right>}(-q^2) \nonumber\\
&+& \hat{L}_M I_{\left<\bar{q}Gq\right>}(-q^2) + \hat{L}_M I_{\left<\bar{q}q\right>^2}(-q^2) \nonumber\\
&+& \left. \hat{L}_M I_{\left<G^3\right>}(-q^2) \right\}, \label{sr}
\end{eqnarray}
where $s_0^D$ is the continuous threshold parameter, $\hat{L}_M$ is the Borel transformation operator. For convenience, we present the expressions for every term in the sum rules (\ref{sr}) in the Appendix.

\section{numerical analysis}

\subsection{Input parameters}

To determine the moments of the $D$-meson leading-twist DA, we take \cite{PDG_Olive:2016xmw}
\begin{eqnarray}
m_{D^-} &=& 1869.59 \pm 0.09 {\rm MeV}, \nonumber\\
\bar{m}_{c}(\bar{m}_c) &=& 1.28 \pm 0.03 {\rm GeV}, \nonumber\\
\bar{m}_d(2{\rm GeV}) &=& 4.7^{+0.5}_{-0.4} {\rm MeV},
\end{eqnarray}
and \cite{BHL_Zhong:2014fma, SRREV_Colangelo:2000dp}
\begin{eqnarray}
\left<\bar{q}q\right>(1 {\rm GeV}) &=& -(240 \pm 10 {\rm MeV})^3, \nonumber\\
\left<\alpha_sG^2\right> &=& 0.038 \pm 0.011 {\rm GeV}^4, \nonumber\\
\left<g_s^3fG^3\right> &=& 0.013 \pm 0.007 {\rm GeV}^6, \nonumber\\
\left<g_s\bar{q}\sigma TGq\right> &=& 0.8 \left<\bar{q}q\right>, \nonumber\\
\left<g_s\bar{q}q\right>^2 &=& 1.8 \times 10^{-3} {\rm GeV}^6.
\end{eqnarray}
The parameters can be run to any other scales by using the renormalization group equation, such as \cite{RGE_Yang:1993bp, RGE_Hwang:1994vp}
\begin{eqnarray}
\bar{m}_c(\mu) &=& \bar{m}_c(\bar{m}_c) \left[ \frac{\alpha_s(\mu)}{\alpha_s(\bar{m}_c)} \right]^{\frac{12}{25}}, \nonumber\\
\bar{m}_d(\mu) &=& \bar{m}_d(2{\rm GeV}) \left[ \frac{\alpha_s(\mu)}{\alpha_s(2{\rm GeV})} \right]^{\frac{12}{27}}, \nonumber\\
\left<\bar{q}q\right>(\mu) &=& \left<\bar{q}q\right>(1{\rm GeV}) \left[ \frac{\alpha_s(\mu)}{\alpha_s(1{\rm GeV})} \right]^{-\frac{12}{27}}.
\end{eqnarray}
The gluon-condensates $\left<\alpha_sG^2\right>$ and $\left<g_s^3fG^3\right>$ are scale-independent, and we ignore the scale-dependence of the four-quark condensate $\left<g_s\bar{q}q\right>^2$, whose value is already very small. Generally, we shall take the renormalization scale as the Borel parameter, $\mu=M$, which represents the typical momentum flow of the process.

The $D$-meson decay constant is taken as the PDG value \cite{PDG_Olive:2016xmw}: $f_D = 203.7 \pm 4.7 \pm 0.6 \rm MeV$. For the continuous threshold $s_0^{D}$, it is usually taken as the square of $D$-meson's first exciting state. Different from the cases of pion and kaon, the $D$-meson's first exciting state has not been experimentally confirmed yet. According to the helicity analysis of Refs.\cite{s0_delAmoSanchez:2010vq, s0_Aaij:2013sza}, Ref.\cite{s0_delAmoSanchez:2010vq} suggests the quantum state of $D^0(2550)$ is $J^P = 0^-$, which has the same quantum number as $D$-meson, e.g., $I(J^P) = \frac{1}{2}(0^-)$. On the other hand, with an sum rules prediction within HQET \cite{s0_Huang:1998sa}, the authors of Ref.\cite{MS_Li:2012gr} suggest $s_0^D = (6.5 \pm 0.25) {\rm GeV}^2$. Thus in this work, we approximately take $D^0(2550)$ as the first excitation state of $D$-meson as suggested by Ref.\cite{s0_delAmoSanchez:2010vq}, and the continuous threshold value is taken as $s_0^D = 6.5025 \textrm{GeV}^2$.

\subsection{The moments $\left<\xi^n\right>_D$ of $\phi_{2;D}$}

\begin{table}[htb]
\caption{Criteria for determining the Borel windows of the $D$-meson leading-twist DA moments $\left<\xi^n\right>_D$. }
\begin{tabular}{ c | c | c }
\hline
~$n$~ & ~continue contribution~ & ~Dimension-six Contribution~ \\
\hline
~$1$~ & ~$<15\%$~ & ~$<1\%$~ \\
~$2$~ & ~$<25\%$~ & ~$<10\%$~ \\
~$3$~ & ~$<20\%$~ & ~$<5\%$~ \\
~$4$~ & ~$<40\%$~ & ~$<15\%$~ \\
\hline
\end{tabular}
\label{tcriterion}
\end{table}

\begin{table}[htb]
\caption{The determined Borel windows and the corresponding $D$-meson leading-twist DA moments $\left<\xi^n\right>_D (n=1,2,3,4)$. All input parameters are set to be their central values. }
\begin{tabular}{ c | c | c }
\hline
~$n$~ & ~$M^2$~ & ~$\left<\xi^n\right>_D$~ \\
\hline
~$1$~ & ~$[3.247,7.035]$~ & ~$[-0.417,-0.397]$~ \\
~$2$~ & ~$[1.862,2.917]$~ & ~$[0.290,0.303]$~ \\
~$3$~ & ~$[3.157,5.763]$~ & ~$[-0.181,-0.175]$~ \\
~$4$~ & ~$[2.410,4.572]$~ & ~$[0.151,0.141]$~ \\
\hline
\end{tabular}
\label{tbw}
\end{table}

To fix the Borel window, one usually requires the most uncertain contributions from both the continuum states and the highest dimensional condensates be a reasonably small value and the sum rules be insensitive to the Borel parameter $M$. The contributions from continuum states and dimension-six condensates dominate the systematic error of the predicted moments $\left<\xi^n\right>_D$, so smaller magnitudes of them indicate better accuracy of the sum rules. In usual treatment, the continuum contribution is taken to be less than $30\%$ and the contribution from dimension-six condensate is less than $10\%$. For the present case, our criteria for the continuum states and the dimension-six condensates contributions are presented in Table \ref{tcriterion}. Table \ref{tcriterion} shows better accuracy of $\left<\xi^1\right>_D$, $\left<\xi^2\right>_D$ and $\left<\xi^3\right>_D$ can be achieved than the usual criteria. In order to obtain the Borel window of $\left<\xi^4\right>_D$, we soften the continuum contribution to be $40\%$, which inversely could lead to lower accuracy for $\left<\xi^4\right>_D$. The determined Borel windows and the corresponding $D$-meson leading-twist DA moments $\left<\xi^n\right>_D$ $(n=1,2,3,4)$ are presented in Table \ref{tbw}, where all input parameters are taken to be their central values.

\begin{figure}[htb]
\centering
\includegraphics[width=0.45\textwidth]{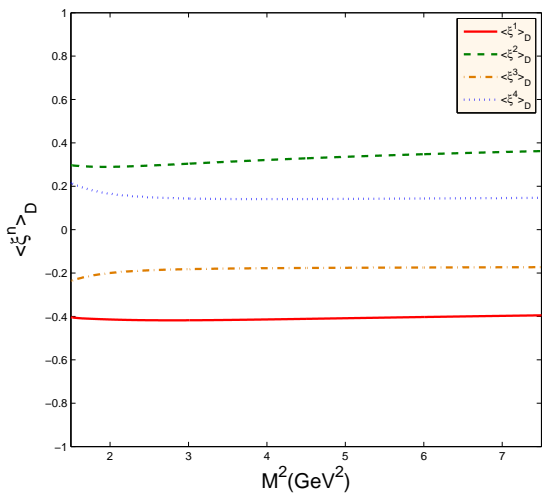}
\caption{The $D$-meson leading-twist DA moments $\left<\xi^n\right>_{D}(n=1,2,3,4)$ versus the Borel parameter $M^2$, where all input parameters are set to be their central values. The solid, dashed, dash-dotted and dotted lines are for $\left<\xi^1\right>_{D}$, $\left<\xi^2\right>_{D}$, $\left<\xi^3\right>_{D}$ and $\left<\xi^4\right>_{D}$, respectively.}
\label{fxinM2}
\end{figure}

\begin{table*}[htb]
\caption{The impact of various inputs on $\left<\xi^n\right>_D$. The Borel parameter $M$ is fixed to be its central value. The labels ``$|_{\rm up}$'' and ``$|_{\rm low}$'' stand for the upper and lower bounds of the inputs, and the symbols ``$+$'' and ``$-$'' represent the positive and negative errors brought by the corresponding input, respectively. }
\begin{tabular}{ c | c c | c c | c c | c c }
\hline
~ & ~$\left<\alpha_sG^2\right>|_{\rm up}$~ & ~$\left<\alpha_sG^2\right>|_{\rm low}$~ & ~$\left<g_s^3fG^3\right>|_{\rm up}$~ & ~$\left<g_s^3fG^3\right>|_{\rm low}$~ & ~$\left<\bar{q}q\right>|_{\rm up}$~ & ~$\left<\bar{q}q\right>|_{\rm low}$~ & ~$\left<g_s\bar{q}\sigma TGq\right>|_{\rm up}$~ & ~$\left<g_s\bar{q}\sigma TGq\right>|_{\rm low}$~ \\
\hline
~$\left<\xi^1\right>_D$~ & ~$-$~ & ~$+$~ & ~$-$~ & ~$+$~ & ~$-$~ & ~$+$~ & ~$+$~ & ~$-$~ \\
~$\left<\xi^2\right>_D$~ & ~$+$~ & ~$-$~ & ~$+$~ & ~$-$~ & ~$+$~ & ~$-$~ & ~$-$~ & ~$+$~ \\
~$\left<\xi^3\right>_D$~ & ~$-$~ & ~$+$~ & ~$-$~ & ~$+$~ & ~$-$~ & ~$+$~ & ~$+$~ & ~$-$~ \\
~$\left<\xi^4\right>_D$~ & ~$+$~ & ~$-$~ & ~$+$~ & ~$-$~ & ~$+$~ & ~$-$~ & ~$-$~ & ~$+$~ \\
\hline\hline
~ & ~$\bar{m}_c|_{\rm up}$~ & ~$\bar{m}_c|_{\rm low}$~ & ~$\bar{m}_d|_{\rm up}$~ & ~$\bar{m}_d|_{\rm low}$~ & ~$m_D|_{\rm up}$~ & ~$m_D|_{\rm low}$~ & ~$f_D|_{\rm up}$~ & ~$f_D|_{\rm low}$~ \\
\hline
~$\left<\xi^1\right>_D$~ & ~$+$~ & ~$+$~ & ~$+$~ & ~$-$~ & ~$-$~ & ~$+$~ & ~$+$~ & ~$-$~ \\
~$\left<\xi^2\right>_D$~ & ~$-$~ & ~$+$~ & ~$-$~ & ~$+$~ & ~$+$~ & ~$-$~ & ~$-$~ & ~$+$~ \\
~$\left<\xi^3\right>_D$~ & ~$+$~ & ~$-$~ & ~$+$~ & ~$-$~ & ~$-$~ & ~$+$~ & ~$+$~ & ~$-$~ \\
~$\left<\xi^4\right>_D$~ & ~$-$~ & ~$+$~ & ~$-$~ & ~$+$~ & ~$+$~ & ~$-$~ & ~$-$~ & ~$+$~ \\
\hline
\end{tabular}
\label{tinputs}
\end{table*}

Fig.\ref{fxinM2} shows the stabilities of the $D$-meson leading-twist DA moments $\left<\xi^n\right>_{D}$ $(n=1,2,3,4)$ in the allowable Borel windows. By taking all uncertainty sources into consideration, we obtain
\begin{eqnarray}
\left<\xi^1\right>_D |_{\mu = 2 \rm GeV} &=& -0.418^{+0.021}_{-0.022}, \nonumber\\
\left<\xi^2\right>_D |_{\mu = 2 \rm GeV} &=& 0.289^{+0.023}_{-0.022}, \nonumber\\
\left<\xi^3\right>_D |_{\mu = 2 \rm GeV} &=& -0.178 \pm 0.010, \nonumber\\
\left<\xi^4\right>_D |_{\mu = 2 \rm GeV} &=& 0.142^{+0.013}_{-0.012},
\label{nxin}
\end{eqnarray}
where the errors are squared averages of all the mentioned error sources. By fixing the Borel parameter $M$ to be its central value of the determined Borel window, Table \ref{tinputs} shows the impact of various inputs on $\left<\xi^n\right>_D$, where the labels ``$|_{\rm up}$'' and ``$|_{\rm low}$'' stand for the upper and lower bounds of the inputs and the symbols ``$+$'' and ``$-$'' represent the positive and negative errors brought by the corresponding inputs, respectively. Table \ref{tinputs} shows that if the upper limit of an input parameter causes a positive error in $\left<\xi^1\right>_D$, it will lead to a positive error for $\left<\xi^3\right>_D$ and lead to negative errors for $\left<\xi^2\right>_D$ and $\left<\xi^4\right>_D$, and vice versa; and if the upper limit of an input parameter leads to a positive error in a moment, its lower bound will lead to a negative error in this moment, and vice versa. The only exception is the $c$-quark current mass $\bar{m}_c$. Fortunately, the error caused by $\bar{m}_c$ is negligible. Thus, it is reasonable to assume that the four moments $\left<\xi^n\right>_D (n=1,2,3,4)$ can not be varied independently, all of which follow the same variation trends as described above. For example, to determine the uncertainty of the leading-twist DA,  if the magnitudes of $\left<\xi^1\right>_D$ and $\left<\xi^3\right>_D$ take the upper bound, the magnitudes of $\left<\xi^2\right>_D$ and $\left<\xi^4\right>_D$ should take the lower bound, and vice versa.

\subsection{The improved Model for the $D$-Meson Leading-Twist DA $\phi_{2;D}$}

\begin{table*}[htb]
\caption{Typical $D$-meson leading-twist DA model parameters at scale $\mu =2{\rm GeV}$.}
\begin{tabular}{ c c c c | c c c c c c}
\hline
~$a_1^D$~& ~$a_2^D$~ & ~$a_3^D$~ & ~$a_4^D$~& ~$A_D({\rm GeV}^{-1})$~ & ~$B^D_1$~ & ~$B^D_2$~& ~$B^D_3$~ & ~$B^D_4$~ & ~$\beta_D({\rm GeV})$~ \\
\hline
~$-0.697$~& ~$0.258$~ & ~$0.009$~ & ~$-0.024$~& ~$1.855$~ & ~$-0.567$~ & ~$0.027$~& ~$0.165$~ & ~$-0.078$~ & ~$5.776$~ \\
~$-0.697^{+0.036}$~& ~$0.258_{-0.064}$~ & ~$0.009^{+0.003}$~ & ~$-0.024_{+0.020}$~& ~$1.909$~ & ~$-0.524$~ & ~$-0.030$~& ~$0.154$~ & ~$-0.049$~ & ~$5.806$~ \\
~$-0.697_{-0.037}$~& ~$0.258^{+0.068}$~ & ~$0.009_{-0.002}$~ & ~$-0.024^{-0.026}$~& ~$1.800$~ & ~$-0.616$~ & ~$0.092$~& ~$0.177$~ & ~$-0.113$~ & ~$5.684$~ \\
\hline
\end{tabular}
\label{DAparameter}
\end{table*}

One can use the DA moments $\left<\xi^n\right>_D$ to get the Gegenbauer moments $a^D_n$. For example, by using the relationship between $\left<\xi^n\right>_{D}$ and  $a_n^D$~\cite{BHL_Zhong:2014fma}, we obtain
\begin{eqnarray}
a_1^D(2 \rm GeV) &=& -0.697^{+0.036}_{-0.037}, \nonumber\\
a_2^D(2 \rm GeV) &=& 0.258^{+0.068}_{-0.064}, \nonumber\\
a_3^D(2 \rm GeV) &=& 0.009^{+0.003}_{-0.002}, \nonumber\\
a_4^D(2 \rm GeV) &=& -0.024_{+0.020}^{-0.026}.
\label{anD}
\end{eqnarray}
Substituting the above Gegenbauer moments $a_n^D$ into Eq.(\ref{an}), together with the constraints (\ref{NC}, \ref{P}), we can determine the input parameters $A_D$, $B_n^D$ and $\beta_D$ for the leading-twist DA $\phi_{2;D}$. The accuracy of $\phi_{2;D}$ is dominated by the accuracy of the Gegenbauer moments $a_D^n$. Table \ref{DAparameter} presents some typical parameters at scale $\mu =2{\rm GeV}$ for typical choices of Gegenbauer moments $a_D^n$. Similar to the case of the DA moments $\left<\xi^n\right>_D$, the Gegenbauer moments $a_n^D$ also can not be varied independently in their own error regions, and the uncertainty of the DA model is determined by the following two sets of $a_n^D$, namely, i) $a_1^D(2\rm GeV) = -0.697^{+0.036}$, $a_2^D(2\rm GeV) = 0.258_{-0.064}$, $a_3^D(2\rm GeV) = 0.009^{+0.003}$, $a_4^D(2\rm GeV) = -0.024_{+0.020}$; ii) $a_1^D(2\rm GeV) = -0.697_{-0.037}$, $a_2^D(2\rm GeV) = 0.258^{+0.068}$, $a_3^D(2\rm GeV) = 0.009_{-0.002}$, $a_4^D(2\rm GeV) = -0.024^{-0.026}$. Table \ref{DAparameter} associates the uncertainty of $\phi_{2;D}$ with the error of Gegenbauer moments $a_D^n$, which facilitates our further discussion on the impact of $\phi_{2;D}$ as an input parameter to the $B\to D$ decay.

\begin{figure}[htb]
\centering
\includegraphics[width=0.45\textwidth]{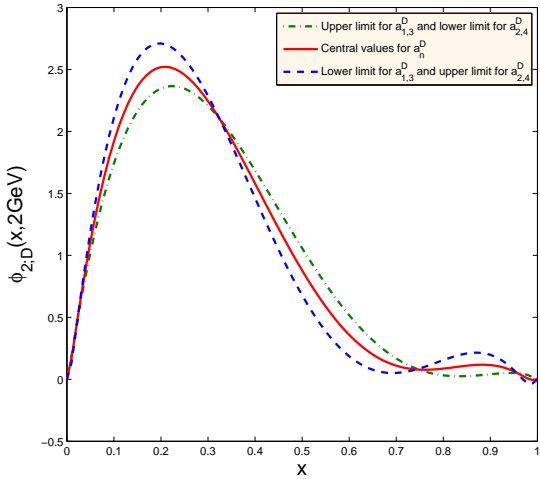}
\caption{The curves of the $D$-meson leading-twist DA $\phi_{2;D}$ with the parameter values exhibited in Table \ref{DAparameter}. }
\label{fDAan}
\end{figure}

\begin{figure}[htb]
\centering
\includegraphics[width=0.45\textwidth]{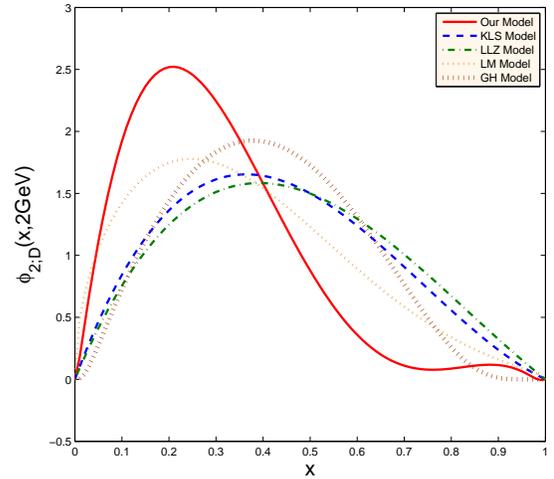}
\caption{A comparison of the $D$-meson leading-twist DA $\phi_{2;D}$. The solid, dashed, dash-dotted, dotted and the thick dotted line are for our present model (\ref{phi}), the KLS model \cite{MODELI_Kurimoto:2002sb, MODELI_Keum:2003js}, the LLZ model \cite{MODELII_Li:2008ts}, the LM model \cite{MODELIII_Li:1999kna}, and the GH model \cite{MOLELIV_Guo:1991eb}, respectively.}
\label{fDAComparision}
\end{figure}

Fig.\ref{fDAan} shows the $D$-meson leading-twist DA $\phi_{2;D}$ with typical values of the input parameters exhibited in Table \ref{DAparameter}. The solid, the dash-dotted and the dashed lines are for the parameters exhibited in second, third and forth lines of Table \ref{DAparameter}. Fig.\ref{fDAComparision} is a comparison of $\phi_{2;D}$, in which the solid, the dashed, the dash-dotted, the dotted and the thick dotted lines are for our present model (\ref{phi}), the Gegenbauer polynomial-like KLS model \cite{MODELI_Kurimoto:2002sb, MODELI_Keum:2003js}, the LLZ model \cite{MODELII_Li:2008ts}, the Gaussian-type LM model \cite{MODELIII_Li:1999kna} and the GH model \cite{MOLELIV_Guo:1991eb}, respectively.  Our model of $\phi_{2;D}$ prefers a narrower behavior in low $x$-region than other models. It has a peak around $x\sim 0.2$, which is consistent with the LM model, but is inconsistent with the KLS, the LLZ, and the GH model which have peaks at a larger $x$ ($\sim 0.3-0.4$).

\begin{figure}[htb]
\centering
\includegraphics[width=0.45\textwidth]{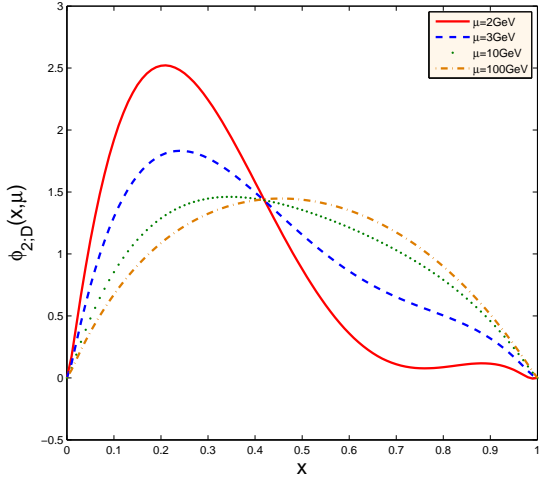}
\caption{The $D$-meson leading-twist DA model (\ref{phi}) at different scales, where the solid, the dashed, the dotted and the dash-dotted lines are for the scales $\mu = 2, 3, 10, 100$ GeV, respectively.}
\label{fDAEverlution}
\end{figure}

Fig.\ref{fDAEverlution} shows the $D$-meson leading-twist DA model (\ref{phi}) at different scales, where the solid, the dashed, the dotted and the dash-dotted lines are for the scales $\mu = 2, 3, 10, 100$ GeV, respectively. It shows that with the increment of $\mu$, $\phi_{2;D}$ becomes broader and broader and becomes more symmetric, e.g. the peak moves closer to $x=0.5$. When $\mu\to\infty$, $\phi_{2;D}$ tends to the well-known asymptotic form, i.e. $\phi_{2;D}(x,\mu\to\infty)=6x(1-x)$.

\subsection{Numerical results of $B\to D$ TFF and its uncertainty from $\phi_{2;D}$}

By using the chiral current correlation function, the LCSR of $f^{B \to D}_{+}$ up to twist-4 accuracy shall involve only the contribution from the $D$-meson leading-twist DA $\phi_{2;D}$ \cite{LCSR_Zuo:2006dk, Zuo:2006re, Fu:2013wqa}. In this subsection, we apply our present DA model to calculate the $B \to D$ TFF $f^{B \to D}_{+}$.

\begin{figure}[htb]
\centering
\includegraphics[width=0.45\textwidth]{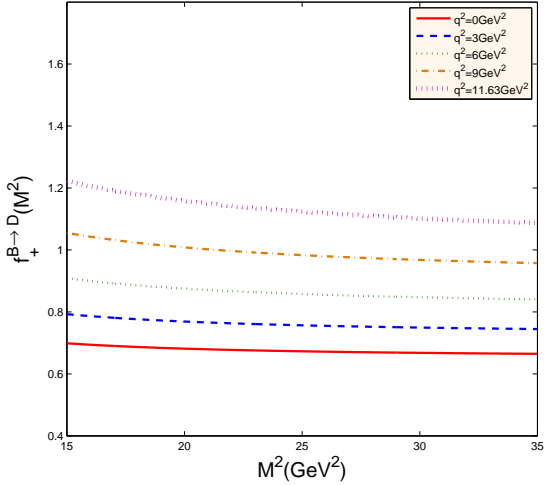}
\caption{The TFF $f^{B\to D}_+(q^2)$ for several $q^2$-values versus the Borel parameter $M^2$.}
\label{ftffm}
\end{figure}

Considering the decay $\overline{B}^0\to D^+l\nu_l$, we take $m_{\overline{B}^0} = 5279.63 \pm 0.15 \rm MeV$ and $\bar{m}_b(\bar{m}_b) = 4.18^{+0.04}_{-0.03} {\rm GeV}$ \cite{PDG_Olive:2016xmw}. For the $B$-meson decay constant, we take the PDG value, $f_B = 188 \pm 17 \pm 18 \rm MeV$ \cite{PDG_Olive:2016xmw}. For the continuum threshold $s_0^B$, we take it to be $s_0^B = 36 \pm 1 \textrm{GeV}^2$. We take the factorization scale to be $\mu \simeq 3 {\rm GeV}$. For the Borel window we take $M^2 = (20-30) \textrm{GeV}^2$. Fig.\ref{ftffm} shows the TFF $f^{B\to D}_+(q^2)$ is stable within the Borel window.

\begin{figure}[htb]
\centering
\includegraphics[width=0.45\textwidth]{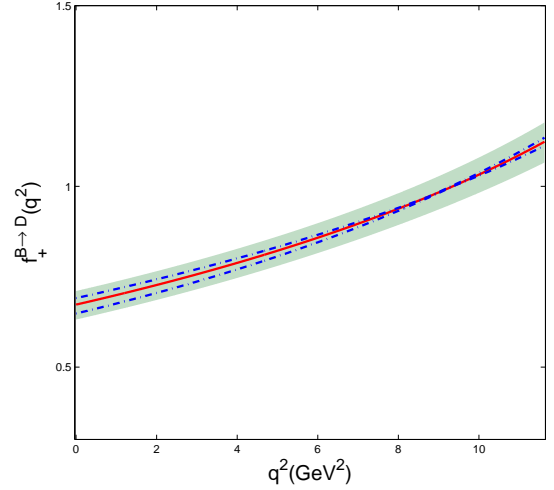}
\caption{The TFF $f^{B\to D}_+(q^2)$ versus $q^2$. The solid line is the central value and the shaded band is the squared average of all the error sources. The error from the leading-twist DA $\phi_{2;D}$ is shown by the dash-dotted lines. }
\label{ftffq}
\end{figure}

We present the TFF $f^{B \to D}_{+}(q^2)$ versus $q^2$ in Fig.\ref{ftffq}. The shaded hand is the theoretical uncertainty, in which the uncertainties from all the mentioned error sources, such as $\phi_{2;D}$, $s_0^B$, $f_{B}$, $f_D$, $m_b$ and etc., have been added up in quadrature. The solid line indicates the central value of $f^{B\to D}_+(q^2)$, the dash-dotted line stands for the uncertainty from DA $\phi_{2;D}$. Fig.\ref{ftffq} shows when $q^2 \in [8,10]{\rm GeV}^2$, the error caused by $\phi_{2;D}$ is rather small, which becomes sizable for $q^2 \in [0,8]{\rm GeV}^2$ and $10{\rm GeV}^2 \leq q^2 \leq q^2_{\rm max} = (m_B - m_D)^2$. This can be numerically explained by the fact that the error of $\phi_{2;D}$  shall be cancelled for the integral region $u \in [\Delta,1]$ of the integral in Eq.(\ref{f+_LCSR}).

Thus in addition to the previously considered error sources, an accurate $\phi_{2;D}$ is also important for achieving a precise $f^{B\to D}_+(q^2)$. For example, at the maximum recoil point with $q^2 = 0$ and the zero recoil point with $q^2 = q^2_{\rm max}$, we have
\begin{widetext}
\begin{eqnarray}
f^{B \to D}_{+}(0) &=& 0.673^{+0.018}_{-0.025}|_{\phi_{2;D}}\ ^{+0.005}_{-0.009}|_{M^2}\ ^{+0.019}_{-0.021}|_{s_0^B} \pm 0.015|_{f_B} \pm 0.016|_{f_D}\ ^{+0.016}_{-0.011}|_{m_b} \nonumber\\
&=& 0.673^{+0.018}_{-0.025}|_{\phi_{2;D}} \pm 0.033|_{\rm Other\ Inputs}
\label{f+0}
\end{eqnarray}
and
\begin{eqnarray}
f^{B \to D}_{+}(q^2_{\rm max}) &=& 1.124 \pm 0.011|_{\phi_{2;D}}\ ^{+0.022}_{-0.035}|_{M^2}\ ^{+0.016}_{-0.018}|_{s_0^B}\ ^{+0.025}_{-0.026}|_{f_B} \pm 0.026|_{f_D}\ ^{+0.027}_{-0.018}|_{m_b} \nonumber\\
&=& 1.124 \pm 0.011|_{\phi_{2;D}}\ ^{+0.052}_{-0.056}|_{\rm Other\ Inputs},
\label{f+max}
\end{eqnarray}
\end{widetext}
where  the error labeled as ``Other Inputs'' is obtained by adding up of all the errors other than the one from $\phi_{2;D}$ in quadrature. The DA $\phi_{2;D}$, the Borel parameter $M^2$, continuum threshold $s_0^B$, $B(D)$-meson decay constant $f_{B(D)}$ and the $b$-quark mass $m_b$ are main error sources. The errors caused by $m_{\overline{B}^0}$ and $m_{D^+}$ are not explicitly shown, because they are less than $10^{-5}$ of the total contributions. Our value in Eq.(\ref{f+0}) agrees with the lattice QCD prediction, $f^{B \to D}_{+}(0) = 0.664 \pm 0.034$~ \cite{LQCD_Na:2015kha}.

 Using the transformation formula $\mathcal{G}(1) = 2\sqrt{m_B m_D}/(m_B + m_D) \times f^{B \to D}_{+}(q^2_{\rm max})$, one can get $\mathcal{G}(1) = 0.987^{+0.047}_{-0.051}$. In the literatures, $\mathcal{G}(1)$ has been calculated with the lattice QCD approach, e.g., $\mathcal{G}(1) = 1.074 \pm 0.018 \pm 0.016$\cite{g1_Hashimoto:1999yp}, $\mathcal{G}(1) = 1.058 \pm 0.016 \pm 0.003^{+0.014}_{-0.005}$\cite{g1_Okamoto:2004xg}, $\mathcal{G}(1) = 1.026 \pm 0.017$\cite{g1_deDivitiis:2007otp}, $\mathcal{G}(1) = 1.0527 \pm 0.0082$\cite{LQCD_Lattice:2015rga} and $\mathcal{G}(1) = 1.035 \pm 0.040$\cite{LQCD_Na:2015kha}. Our result in (\ref{f+max}) is slightly smaller than the values in Refs.\cite{g1_Hashimoto:1999yp, g1_Okamoto:2004xg,  LQCD_Lattice:2015rga}, but is consistent with the values in Ref.\cite{g1_deDivitiis:2007otp,LQCD_Na:2015kha} within reasonable errors.

Furthermore, one can calculate the branching ratio $\mathcal{B}(B\to Dl\bar{\nu}_l)$ with the following two formulas,
\begin{eqnarray} &&
\frac{d}{dq^2} \Gamma(B\to Dl\bar{\nu}_l) \nonumber\\
&& \quad\quad = \frac{G_F^2 |V_{\rm cb}|^2}{192\pi^3 m_B^3} \lambda^{3/2}(q^2) |f_+^{B\to D}(q^2)|^2,
\label{Dif_Dec_Wid}
\end{eqnarray}
\begin{eqnarray}
\mathcal{B}(B\to Dl\bar{\nu}_l) = \tau_B \int^{(m_B - m_D)^2}_{0} dq^2 \frac{d\Gamma(B\to Dl\bar{\nu}_l)}{dq^2},
\label{Bra_Rat}
\end{eqnarray}
where $\lambda(q^2) = (m_B^2 + m_D^2 - q^2)^2 - 4m_B^2 m_D^2$ is the phase-space factor. We take the Fermi constant $G_F = 1.1663787(6) \times 10^{-5} GeV^{-2}$, the $B$ meson lifetime $\tau_{B} = (1.520 \pm 0.004) \times 10^{-12} s$ and the CKM matrix element $|V_{\rm cb}| = (40.5 \pm 1.5) \times 10^{-3}$\cite{PDG_Olive:2016xmw}. Then
\begin{eqnarray}
\mathcal{B}(\overline{B}^0\to D^+l\nu_l) &=& \left(2.132 \pm 0.273\right) \times 10^{-2}.
\label{Br}
\end{eqnarray}
Our $\mathcal{B}(\overline{B}^0\to D^+l\nu_l)$ in (\ref{Br}) agrees with $\mathcal{B}(\overline{B}^0\to D^+l\nu_l) = 2.03^{+0.92}_{-0.70}$ by pQCD \cite{PQCD_Fan:2013qz}, $\mathcal{B}(\overline{B}^0\to D^+l\nu_l) = 2.13^{+0.19}_{-0.18}$ by HQET \cite{HQET_Fajfer:2012vx} and $\mathcal{B}(\overline{B}^0\to D^+l\nu_l) = 2.18 \pm 0.12$ in PDG \cite{PDG_Olive:2016xmw}.

\section{summary}

In this paper, we have made a detailed study on the DA $\phi_{2;D}$ with the QCD sum rules under the framework of the background field theory and tried to estimate the uncertainty from the improved model distribution amplitude. In order to get more accuracy information on the DA $\phi_{2;D}$, we calculate the first four moments $\left<\xi^n\right>_D$ of $\phi_{2;D}$ with QCD sum rules in the framework of BFT. Their values are obtained as: $\left<\xi^1\right>_D = -0.418^{+0.021}_{-0.022}$, $\left<\xi^2\right>_D = 0.289^{+0.023}_{-0.022}$, $\left<\xi^3\right>_D = -0.178 \pm 0.010$ and $\left<\xi^4\right>_D = 0.142^{+0.013}_{-0.012}$ at scale $\mu = 2 \rm GeV$. Furthermore, under the same scale the Gegenbauer moments of $\phi_{2;D}$ are obtained as $a_1^D = -0.697^{+0.036}_{-0.037}$, $a_2^D = 0.258^{+0.068}_{-0.064}$, $a_3^D = 0.009^{+0.003}_{-0.002}$, $a_4^D = -0.024_{+0.020}^{-0.026}$. Based on those four Gegenbauer moments, the improved model for the $D$-meson leading-twist DA $\phi_{2;D}$ has been constructed. Our model has a narrower form than the models existed in the literature, whose peak is at about $x\sim 0.2$. We have also analyzed the effect of $a^n_D$'s uncertainty on the DA $\phi_{2;D}$, which helps us to discuss the uncertainty that occurs when the $\phi_{2;D}$ is used as an input parameter to the exclusive processes.

With our model of $\phi_{2;D}$, we calculate the $B\to D$ TFF $f^{B\to D}_+(q^2)$, and obtain $f^{B\to D}_+(0) = 0.673^{+0.038}_{-0.041}$ and $f^{B\to D}_+(q^2_{\rm max}) = 1.124^{+0.053}_{-0.058}$, we find that the error brought by $\phi_{2;D}$ to $f^{B\to D}_+(q^2)$ is obvious in the low and intermediate $q^2$-region. This case shows that it is very necessary to study and find more accurate form of the meson DA. In the study of various processes, the error caused by the meson DA as an input parameter should be taken into account. Furthermore, we obtain the branching ratio $\mathcal{B}(\overline{B}^0\to D^+l\nu_l) = \left(2.132 \pm 0.273\right) \times 10^{-2}$, which is consistent with experimental data and other approaches in the error range.  \\

{\bf Acknowledgments}:
This work was supported in part by the Natural Science Foundation of China under Grant No.11547015, No.11625520 , No.11575110, No.11405047, No.11765007 and No.11647112.

\appendix
\section{The formulas of those terms in the sum rules (\ref{sr})}

\begin{widetext}
The formulas of those terms in sum rules (\ref{sr}) are
\begin{eqnarray}
{\rm Im} I_{\rm pert}(s) &=& \frac{3}{8\pi (n+1)(n+3)} \left\{ \left[ 2(n+1) \frac{m_c^2}{s} \left( 1 - \frac{m_c^2}{s} \right) + 1 \right] \left( 1 - \frac{2m_c^2}{s} \right)^{n+1} + (-1)^n \right\}, \label{Impert} \\
\hat{L}_M I_{\left<\bar{q}q\right>}(-q^2) &=& (-1)^n \exp \left[ - \frac{m_c^2}{M^2} \right] \frac{m_q \left< \bar{q}{q} \right>}{M^4}, \label{qq} \\
\hat{L}_M I_{\left<G^2\right>}(-q^2) &=& \frac{\left<\alpha_sG^2\right>}{M^4} \frac{1}{12\pi} \left[ 2n(n-1) \mathcal{H}(n-2,1,1) + \mathcal{H}(n,0,0) - \frac{m_c^2}{M^2} \mathcal{H}(n,1,-2) \right], \label{GG} \\
\hat{L}_M I_{\left<\bar{q}Gq\right>}(-q^2) &=& (-1)^n \exp \left[ - \frac{m_c^2}{M^2} \right] \frac{m_q \left< g_s\bar{q}\sigma TGq \right>}{M^6} \left[ -\frac{8n+1}{18} - \frac{2m_c^2}{9M^2} \right], \label{qGq} \\
\hat{L}_M I_{\left<\bar{q}q\right>^2}(-q^2) &=& (-1)^n \exp \left[ - \frac{m_c^2}{M^2} \right] \frac{\left<g_s\bar{q}{q}\right>^2}{M^6} \frac{2(2n+1)}{81}, \label{qq2} \\
\hat{L}_M I_{\left<G^3\right>}(-q^2) &=& \frac{\left<g_s^3fG^3\right>}{M^6} \exp \left[ -\frac{m_c^2}{M^2} \right] \frac{1}{\pi^2} \left\{ -\frac{17}{96} \mathcal{F}_1(n,5,3,2,\infty) + \frac{n}{144} \mathcal{F}_2(n-1,5,3,1,\infty) - \frac{1}{96} \mathcal{F}_2(n,4,3,1,\infty) \right. \nonumber\\
&+& \frac{1}{144} \mathcal{F}_2(n,3,3,1,\infty) - \frac{17}{96} \mathcal{G}_1(n,5) - \frac{17}{32} \mathcal{G}_2(n,5) \left( 1 - \frac{1}{3} \frac{m_c^2}{M^2} \right) + \frac{n}{144} \mathcal{G}_2(n-1,5) - \frac{n}{96} \mathcal{G}_3(n,4) \nonumber\\
&+& \frac{n}{144} \mathcal{G}_3(n,3) + \frac{1}{288} \left[ 204\delta^{n0} + 204\theta(n-1)(-1)^n + (-1)^n \left( 100n - 154 + 51\frac{m_c^2}{M^2} \right) \right] \left[ \ln \frac{M^2}{\mu^2} + \psi(3) \right] \nonumber\\
&+& \left. \frac{(-1)^n}{288} \left( 17\frac{m_c^2}{M^2} - 4n \right) \right\} + \frac{\left<g_s^3fG^3\right>}{M^6} \frac{1}{\pi^2} \left\{ \frac{1}{288} \left[ -4(n+1)n(n-1) \mathcal{H}(n-2,1,1) + 4(n+1) \mathcal{H}(n,0,0) \right.\right. \nonumber\\
&-& \left. 2n \mathcal{H}(n-1,1,-1) - 3 \mathcal{H}(n,0,-1) - 51 \mathcal{H}(n,1,-2) \right] + \frac{1}{288} \frac{m_c^2}{M^2} \left[ -4n(n-1) \mathcal{H}(n-2,1,0) \right. \nonumber\\
&-& \left.\left. 2 \mathcal{H}(n,0,-2) + 4 \mathcal{H}(n,0,-1) - 2 \mathcal{H}(n-1,1,-2) - 3 \mathcal{H}(n,1,-3) \right] + \frac{1}{240} \frac{m_c^4}{M^4} \mathcal{H}(n,1,-4) \right\}, \label{GGG}
\end{eqnarray}
where
\begin{eqnarray}
\mathcal{F}_1(n,a,b,l_{\rm min},l_{\rm max}) &=& \sum_{k=0}^n \frac{(-1)^k n! \Gamma(k+a)}{k!(n-k)!} \sum_{l=l_{\rm min}}^{l_{\rm max}} \frac{\Gamma(l+b) \Gamma(n-1-k+l)}{\Gamma(n-1+l+a)} \sum^l_{i=0} \frac{1}{i! (l-i)! (l-1-i+b)!} \left( -\frac{m_c^2}{M^2} \right)^{l-i}, \label{F1}\\
\mathcal{F}_2(n,a,b,l_{\rm min},l_{\rm max}) &=& \sum_{k=0}^n \frac{(-1)^k n! \Gamma(k+a)}{k!(n-k)!} \sum_{l=l_{\rm min}}^{l_{\rm max}} \frac{\Gamma(l+b) \Gamma(n-k+l)}{\Gamma(n+l+a)} \sum^l_{i=0} \frac{1}{i! (l-i)! (l-1-i+b)!} \left( -\frac{m_c^2}{M^2} \right)^{l-i}, \label{F2}\\
\mathcal{G}_1(n,a) &=& \sum^{n-2}_{k=0} \frac{(-1)^k n! \Gamma(k+a) \Gamma(n-1-k)}{k! (n-k)! \Gamma(n-1+a)}, \label{G1}\\
\mathcal{G}_2(n,a) &=& \sum^{n-1}_{k=0} \frac{(-1)^k n! \Gamma(k+a) \Gamma(n-k)}{k! (n-k)! \Gamma(n+a)}, \label{G2}\\
\mathcal{G}_3(n,a) &=& \sum^{n-1}_{k=0} \frac{(-1)^k (n-1)! \Gamma(k+a) \Gamma(n-k)}{k! (n-k)! \Gamma(n+a)}, \label{G3}\\
\mathcal{H}(n,a,b) &=& \int^1_0 dx (2x-1)^n x^a (1-x)^b \exp \left[ -\frac{m_c^2}{M^2 (1-x)} \right]. \label{H}
\end{eqnarray}
In calculation, the following Borel transformation formulas are adopted,
\begin{eqnarray} &&
\hat{L}_M \frac{1}{(-q^2 + m_c^2)^k} \ln \frac{-q^2 + m_c^2}{\mu^2} = \frac{1}{(k-1)!} \frac{1}{M^{2k}} e^{-\frac{m_c^2}{M^2}} \left[ \ln \frac{M^2}{\mu^2} + \psi(k) \right] \quad (k \geq 1), \nonumber\\ &&
\hat{L}_M (-q^2 + m_c^2)^k \ln \frac{-q^2 + m_c^2}{\mu^2} = (-1)^{k+1} k! M^{2k} e^{-\frac{m_c^2}{M^2}} \quad (k \geq 0), \nonumber\\&&
\hat{L}_M \frac{(-q^2)^l}{(-q^2+m_c^2)^{l+\tau}} = \left\{
\begin{array}{l l}
0, & \tau = 0, l = 0;\\
\sum^{l-1}_{i=0} \frac{l!}{i! (l-i)! (l-i-1)!} \left( -\frac{m_c^2}{M^2} \right)^{l-i} e^{-\frac{m_c^2}{M^2}}, & \tau = 0, l > 0;\\
\sum^{l}_{i=0} \frac{l!}{i! (l-i)! (l+\tau-i-1)!} \left( -\frac{m_c^2}{M^2} \right)^{l-i} \frac{1}{M^{2\tau}} e^{-\frac{m_c^2}{M^2}}, & \tau > 0, l \geq 0.
\end{array}
\right.
\end{eqnarray}
\end{widetext}

\end{document}